\documentclass{ansconf}
\usepackage{graphicx}
\usepackage{tabls}
\usepackage{multirow}
\usepackage{subfig}

\usepackage{fancyhdr,graphicx,lastpage}
\usepackage{float}
\fancypagestyle{plain}{
\fancyhf{}

\fancyfoot[C]{\thepage}
}

\begin{document}

\title{TOWARD DIRECT NUMERICAL SIMULATION OF TURBULENT AND TRANSITIONAL FLOW IN HEXAGONAL SUBCHANNELS FOR HELIUM CONDITIONS}

\author{Carolina S. B. Dutra}
\author{Elia Merzari}
\affil{Ken and Mary Alice Lindquist Department of Nuclear Engineering \\
Pennsylvania State University\\
  University Park, PA 16802 \\
  cpd5488@psu.edu; ebm5351@psu.edu }

\maketitle

\begin{abstract}
The understanding of the coolant thermal hydraulics in rod bundles is essential to the design of nuclear reactors. However, flows with low Reynolds numbers present serious modeling challenges especially in the presence of heat transfer and natural convection. They are notoriously difficult to analyze through standard Computational Fluid Dynamics (CFD) tools. High fidelity simulations, such as the Direct Numerical Simulations (DNS), can provide invaluable insight into the flow physics, supporting experiments in developing a deeper understanding and eventually enabling the accurate simulation of this class of flows. In fact, data generated from these high fidelity methods can then be used to benchmark available turbulence models and deliver cheap faster running methods.
\par In the present work, the convective heat transfer in hexagonal subchannels was studied through a DNS approach, using the high-order spectral element method code Nek5000, developed at Argonne National Laboratory. First, the geometric model composed of two hexagonal arrayed rod bundle subchannels with a pitch-to-diameter ratio of 1.5 is built, and then, the mesh is generated. This unusually high P/D, combined with low Reynolds numbers represent  conditions of interest for Gas fast reactors (GFRs). To our knowledge there is no available dataset at these conditions.  

In this work we detailed the development of the numerical benchmark and a series of preliminary LES simulations.
Periodic boundary conditions  are applied in the streamwise and spanwise directions. Non-slip boundary conditions are applied  at the wall. Four different cases are studied, with Reynolds of Re=2500, 5000, 7500 and 10000. All calculations have been performed with the Prandtl number of 0.61, corresponding to helium conditions of interest for Gas Fast Reactor applications. The results are analyzed with a polynomial-order convergence study, and the Reynolds-stresses and the turbulent kinetic budgets are presented and discussed.

\raggedleft
\end{abstract}

\raggedright

\section{Introduction}
\par Understanding the flow behavior and the heat transfer mechanisms in rod bundles for a range of coolant types is critical to developing advanced reactor designs. A broad range of designs relies on rod bundles as the geometric configuration of the core, including gas fast reactors being proposed in the United States \cite{Choi2017}. Such designs, cooled by helium and using a hexagonal lattice with a relatively high Pitch to Diameter ratio (P/D), are the focus of this study.
\par Most of the nuclear thermal-hydraulic problems are inherently multiscale. As the new designs of reactors lack the extensive integral testing of LWRs, their accurate modeling requires corresponding practical approaches that combine separate effect testing with high fidelity simulation \cite{Tarantino2020}. Despite the tremendous growth of computational power in the past decades, modeling thermal fluid problems for engineering with a direct resolution of all scales is still too computationally expensive. 
\par As a result, a computational analyst has to filter micro-scale information. This leads to scaling gaps associated with missing physics and corresponding epistemic uncertainties. Closure models are needed to address the scaling gaps and correctly account for the micro-scale physics. However, the attempts to derive generic closures for many problems were unsuccessful due to its high dimensionality and nonlinearity. As a result, most of those models are designed to work in specific flow conditions.
\par Modeling heat transfer in a nuclear reactor is an example of a problem involving multiple spatial and temporal scales (from the interaction of small (Kolmogorov scale) and large turbulent eddies to the influence of the reactor geometry). Flexible closure models that would correctly account for the effects of turbulence in a variety of flow conditions are a challenging problem. This includes both the development of adequate Reynolds-Averaged Navier-Stokes (RANS) models in computational fluid dynamics on the one hand and the development of proper heat transfer correlations for systems codes on the other. 

\par RANS models have been proven to provide reasonable results for simple flows under single-phase conditions \cite{Baglietto2005} \cite{Conner2010}. However, they cannot accurately predict the flow in rod bundles as they differ from the pipe and parallel channel flows and present a complex behavior \cite{Merzari2011} \cite{Merzari2017} \cite{Merzari2020}. Therefore, the high-fidelity approaches are needed to provide a detailed flow characterization to validate lower-fidelity ones, modeling larger domains. For that reason, we need to use Large Scale Simulations (LES) and Direct Numerical Simulations (DNS) to resolve the small scales of turbulence in a single rod arrangement and provide accurate heat transfer calculations \cite{Tiselj2020} to derive into closure models. 

\par From the aspect of DNS, the uneven resolution requirement for turbulent temperature and velocity fields challenges the efficient utilization of computational resources \cite{Merzari2020a}.
The dissimilarity between turbulent momentum and heat transfer is observed in the DNS study on the low Reynolds forced convection in the prototypic planar channel \cite{Kawamura1998}. The temperature field showed conduction-dominated distribution, whereas the turbulent behavior of the velocity field can be seen from the near-wall profile. Further in the prototypic mixed convection case, Poiseuille-Rayleigh-Bénard (PRB) flow, the presence of a large-scale convective structure significantly altered the behavior of turbulent momentum and heat transfer. Enhancement of turbulent mixing in the channel core and the wall-normal turbulent heat flux can be observed \cite{Guo2020} \cite{Santis2018}.

\par In this work, we aim to develop turbulent flow and heat transfer models for ceramic-clad fuel cores with helium coolant during normal and shutdown conditions, to provide high-fidelity numerical data to validate multi-scale approaches for new reactors.  

\par  We use the open-source spectral-element code Nek5000 \cite{nek5000}, which solves the incompressible Navier-Stokes equations to model the heat transfer in a triangular lattice subchannel unit cell of a Gas Fast Reactor. Fuel pin bundle calculations from different CFD commercial codes were in agreement with the ones from Nek5000 \cite{Merzari2016}. In particular, model a hexagonal subchannel, with Helium as a coolant under low flow conditions, with a Prandtl number of 0.61 and Reynolds numbers below 10000.

\par Firstly, we examine resolution requirements for DNS simulations by performing RANS simulations. Such calculations allow us to determine the Kolgomorov length scale. We also vary the polynomial orders and the mesh element size to confirm mesh convergence of the RANS results. This enabled us to generate a  mesh that accurately represents the Kolgomorov length scales. 

\par As a second step, we perform a series of LES simulations for the same geometry. These simulations allow us to examine some salient flow features in this geometry. Furthermore, we extract the field data to compare the RANS results with the averaged LES results, confirming the validity of the initial RANS guess. In the future,  we aim to perform the Direct Numerical Simulations (DNS) simulations to obtain the turbulent kinetic budgets and Reynolds-stresses.

\section{Methodology}

The RANS, LES, and the DNS calculations are performed with the open-source spectral-element CFD code, Nek5000 \cite{nek5000}, developed at the Argonne National Laboratory. The domain, composed of hexahedral elements, is discretized with piecewise polynomial functions to solve the incompressible Navier-Stokes equations. The solution is represented as a tensor-product of Nth-order Lagrange polynomials $P_N$ based on Gauss-Lobatto-Legendre (GLL) quadrature. 

\par In DNS, the governing equations are given by the Momentum, Continuity and Energy Equations (\ref{eq1} - \ref{eq3}).

\begin{equation}
    \frac{\partial u_i}{\partial t} + \frac{\partial (u_i u_j)}{\partial x_j} = -\frac{1}{\rho} \frac{\partial p}{\partial x_i} + \frac{\partial}{\partial x_j} \left [ \nu \left ( \frac{\partial u_i}{\partial x_j} + \frac{\partial u_j}{\partial x_i}\right ) \right ]
    \label{eq1}
\end{equation}

\begin{equation}
    \frac{\partial u_i}{\partial x_i} = 0
    \label{eq2}
\end{equation}

\begin{equation}
    \rho C_p \left ( \frac{\partial T}{\partial t} + u_j\frac{\partial T}{\partial x_j}\right ) = \frac{\partial}{\partial x_j} \left( k \frac{\partial T}{\partial x_j}\right )
    \label{eq3}
\end{equation}

where $u$ is the velocity, $t$ is the time, $\rho$ is the density, $\nu$ is the kinematic viscosity, $T$ is the temperature, $C_p$ is the heat capacity, $\lambda$ is the thermal conductivity, and $i$ and $j$ denotes the spacial coordinates.

\subsection{Reynolds-Averaged Navier-Stokes}

To perform the preliminary RANS simulations, the $k-\tau$ turbulence model was chosen due to its robustness and capability to avoid the numerical discretization errors that could occur with other models \cite{Kok2000}.

\par The governing equations of this model are given by:
\begin{equation}
    \frac{\partial \rho k}{\partial t} + \bigtriangledown \cdot (\rho k \vec{u}) = P_k - \beta_{k} \rho \omega k +\bigtriangledown \cdot (\mu_k \bigtriangledown k)
\end{equation}
\begin{equation}
\frac{\partial \rho \tau}{\partial t} + \bigtriangledown \cdot (\rho \tau \vec{u}) = -P_{\tau}v+ \beta_{\omega} \rho +\bigtriangledown \dot (\mu_{\omega} \bigtriangledown\tau) - 8\mu_{\omega} \left \| \bigtriangledown \sqrt{\tau} \right \|^2 + C_D
\end{equation}
\begin{equation}
    P_\tau = \frac{\alpha_{\omega }\tau}{k} P_k,   C_D = \delta_{d} \rho \tau \cdot min \{\bigtriangledown k \cdot \bigtriangledown \tau, 0\}
\end{equation}
where $k$ is the turbulent kinetic energy, $\rho$ is the density, $\vec{u}$ is the velocity, $\mu$ is the diffusion coefficient, $P$ is the production term, $C_D$ is the cross diffusion term and $\tau$ is the inverse of the specific dissipation $\omega$, given by:
\begin{equation}
    \tau = \frac{1}{\omega} = \frac{k}{\epsilon}
\end{equation}
with $\epsilon$ being the turbulent dissipation rate.

\par Later, other RANS models - the $k-\omega $SST and the low-Re $k-\omega$ \cite{Menter1996} - were used and compared to the results.

\subsection{Large Eddy Simulations}
While in the RANS approach, the turbulence is treated as time-averaged and is modeled through differential equations, the LES resolves the largest turbulence scales and models the smallest scales. With the use of a high-performance computer, this method is suitable to model flows with low Reynolds numbers and small domains. 

\par In fact, in LES the filtered version of the Navier-Stokes equations is solved. The subgrid model used is based on an explicit spatial filter \cite{fischer2001filter} that mimics the effect of dissipation associated with scales below the inertial range. This approach is typically standard for high-order methods with minimal dissipation. In this work, in particular, an explicit filter of 2\% is used to remove energy in the highest wavenumber mode within each element. 

\par The results obtained from the LES post-processing are utilized to compute the turbulent kinetic energy $k$:
\begin{equation}
    k = \frac{1}{2}(\overline{u'u'}+\overline{v'v'}+\overline{w'w'})
    \label{eq9}
\end{equation}
where $\overline{u'u'}$, $\overline{v'v'}$ and $\overline{w'w'}$ are the diagonal components of the Reynolds stress tensor.

\subsection{Direct Numerical Simulations}
Albeit the LES can capture the smaller components of the turbulent fields and improve the models for the RANS simulations, only the DNS can capture all the scales of turbulence, as they do not employ models to describe the turbulence when solving the Navier-Stokes set of equations. The DNS resolves all the relevant flow spatial and temporal scales, and therefore it can be considered a true numerical experiment.

\par However, a mesh sensitivity study must be conducted to ensure that the grid is fine enough to capture the smallest eddies down to the Kolgomorov length scale.

\subsection{Case setup} 
The simulations were conducted for heat transfer cases under low-flow conditions, with Reynolds number below 10000, which is defined as:
\begin{equation}
    Re = \frac{\rho v D_h}{\nu},
\end{equation}
where $D_h$ is the hydraulic diameter.
\\In that way, we can obtain insight into the flow field over different conditions, from transitional to turbulent. The coolant is constituted by helium, with a Prandtl number of 0.61.

\par
\begin{figure}[h]
  \centering
     \includegraphics[width=0.63\textwidth]{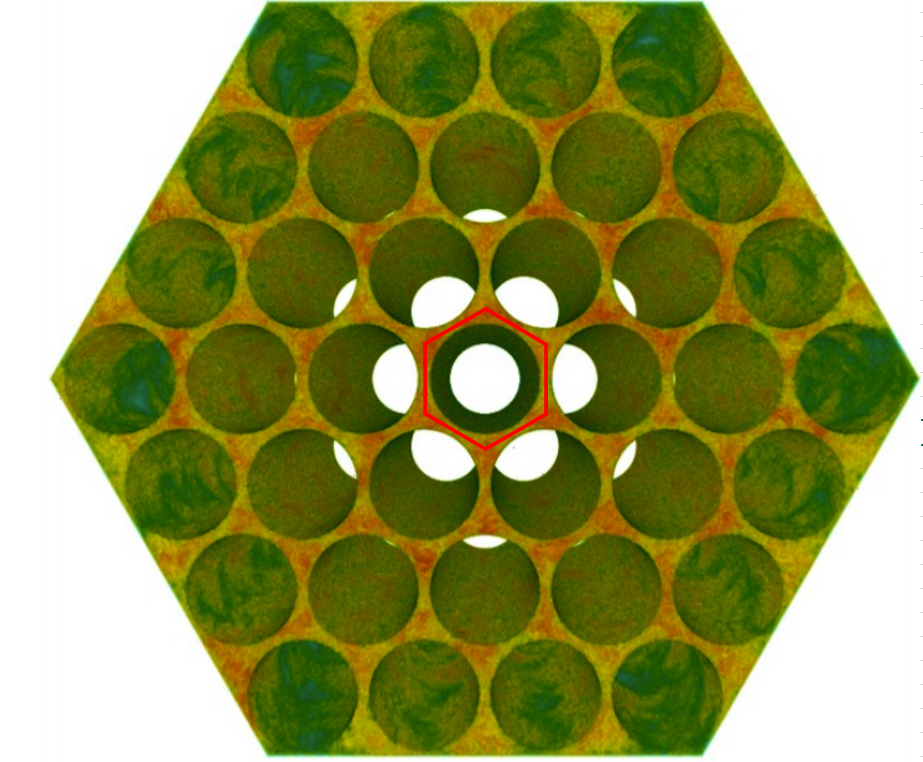}
  \caption{Example of the triangular lattice subchannel simulation \cite{Merzari2016}}
  \label{fig1}
\end{figure}

\begin{figure}[h]
  \centering
     \includegraphics[width=0.57\textwidth]{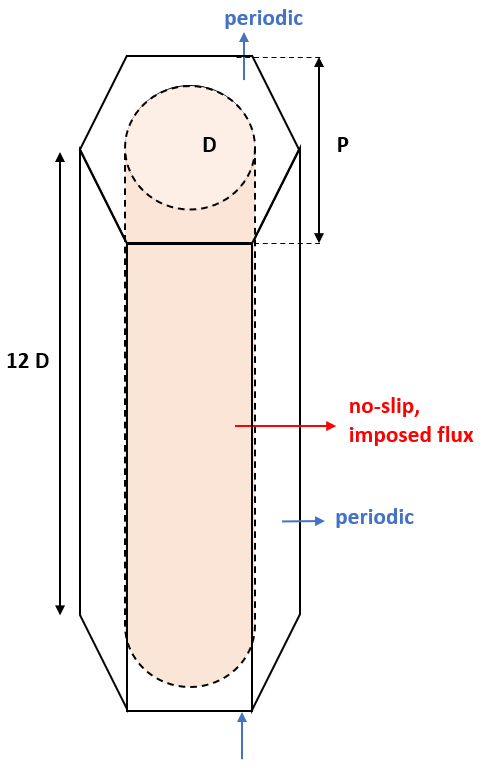}
  \caption{Physical model and Boundary Conditions}
  \label{fig2}
\end{figure}

 The case studied in this work is of a triangular lattice subchannel unit cell, as represented in Figure \ref{fig1}, with periodic temperature and velocity boundary conditions applied to the streamwise direction and the cross-section. This simplified configuration was chosen to save computational cost \cite{merzari2015}. There is a prescribed heat flux at the wall, as shown in Figure \ref{fig2}. The large pitch-to-diameter ratio of 1.5 was chosen due to its relevance to GFR conditions. The periodic streamwise length is chosen to be $12D$, through a series of two-point correlation tests that demonstrated to be sufficient to reach a lack of correlation between the midpoint and the domain ends. 

\par To reduce the problem as periodic for the temperature, we assume the temperature source as constant \cite{Merzari2012}. We then perform a transformation of variables to obtain the Equation \ref{eq4} for the temperature, which represents a linear gradient in the streamwise direction superposed to a periodic temperature field $\tilde{T}$.
 \begin{equation}
     T = T_0 + \gamma z+ \tilde{T}
     \label{eq4}
 \end{equation}
 By conservation of energy we obtain a sink term:
 \begin{equation}
     \gamma = q \frac{C_{p,s}}{C_{p,f}} \frac{A_s}{A_f} \frac{1}{W_{bulk}}
      \label{eq5}
 \end{equation}
where $A_s/A_f$ is the ratio between the area of the solid and the fluid, on the plane normal to the streamwise direction z, $W_{bulk}$ is the streamwise bulk velocity, and $q$ is the heat source. And $\gamma$ depends on the ratio between the heat capacity of the solid and heat capacity of the fluid, $C_{p,s}$ and $C_{p,f}$. Then we substitute Equation \ref{eq4} in Equation \ref{eq3} to obtain a new equation temperature in the fluid:
\begin{equation}
    \frac{\partial \tilde{T}}{\partial t} + u_j \frac{\partial \tilde{T}}{\partial x_j} = \frac{\lambda}{\rho C_p}  \frac{\partial ^2 \tilde{T}}{\partial x_j\partial x_j}  - u_3 \gamma
    \label{eq6}
\end{equation}

\section{Results}

In this section we discuss the results of the analysis conducted in the present work.

\subsection{Mesh sensitivity study}

\par For fluids where the Prandtl number is below unity due to larger thermal diffusivity over kinematic viscosity, the Kolmogorov length scale $\eta$ is the limiting criterion for spatial resolution.

\par In order to verify if the mesh is sufficient to represent the eddies, we use the turbulent kinetic energy $k$ obtained from the RANS simulations to compute the Kolmogorov lengthscale $\eta$ throughout all the domain   \cite{Grotzbach1999}:
\begin{equation}
    \eta = \left (  \frac{\nu}{\epsilon} ^3\right )^{1/4}
    \label{eq7}
\end{equation}

\par
\begin{figure}[H]
  \centering
     \includegraphics[width=0.96 \textwidth]{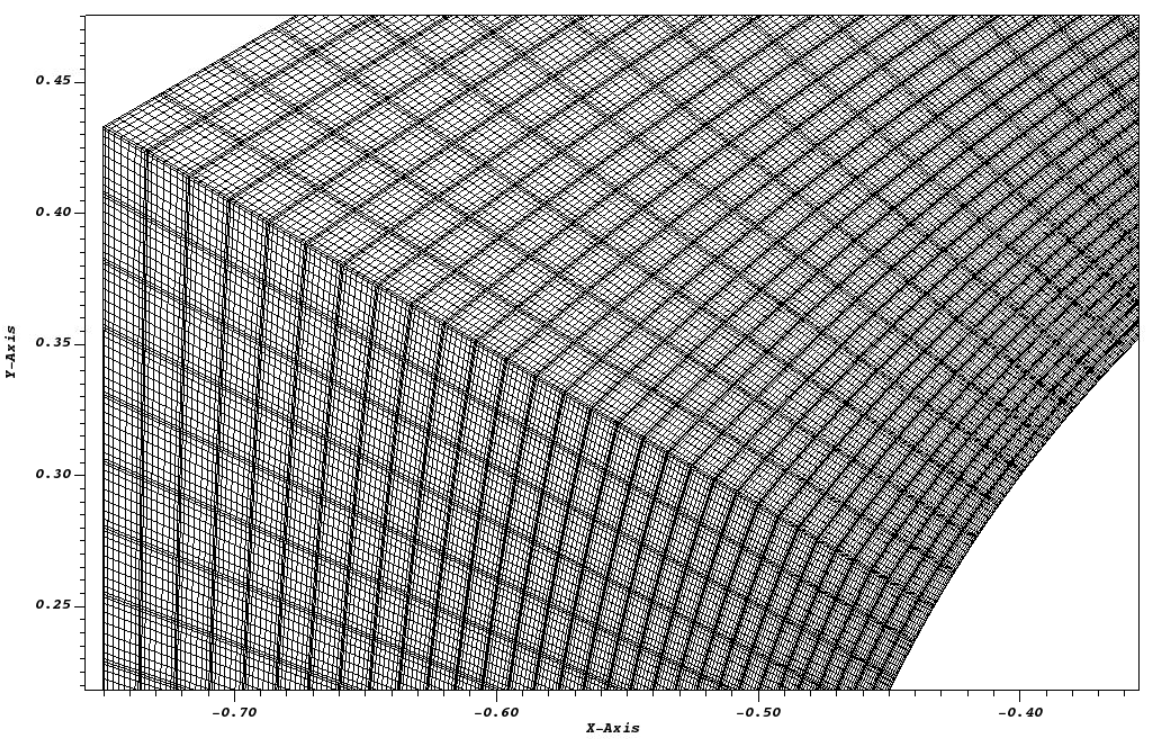}
  \caption{Mesh detail of the cross-section of the domain}
  \label{fig3}
\end{figure}

 A sensitivity analysis was performed to obtain an optimum structured mesh with hexahedral elements to conduct the DNS. The Gauss-Lobatto Legendre (GLL) orders were also varied to satisfy the target grid resolution. The GLL of 11 was applied to obtain a mesh with 1.4 M elements. A cut of the mesh is represented in Figure \ref{fig3}. 

\begin{figure}[h]%
    \centering
    \subfloat \centering {{\includegraphics[width=8cm]{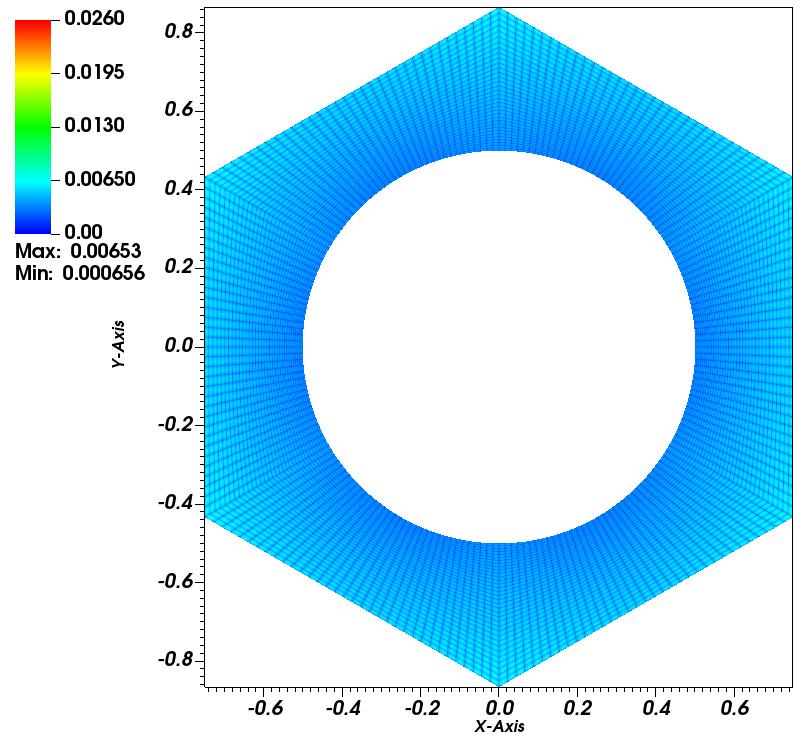} }}%
    \subfloat \centering {{\includegraphics[width=8cm]{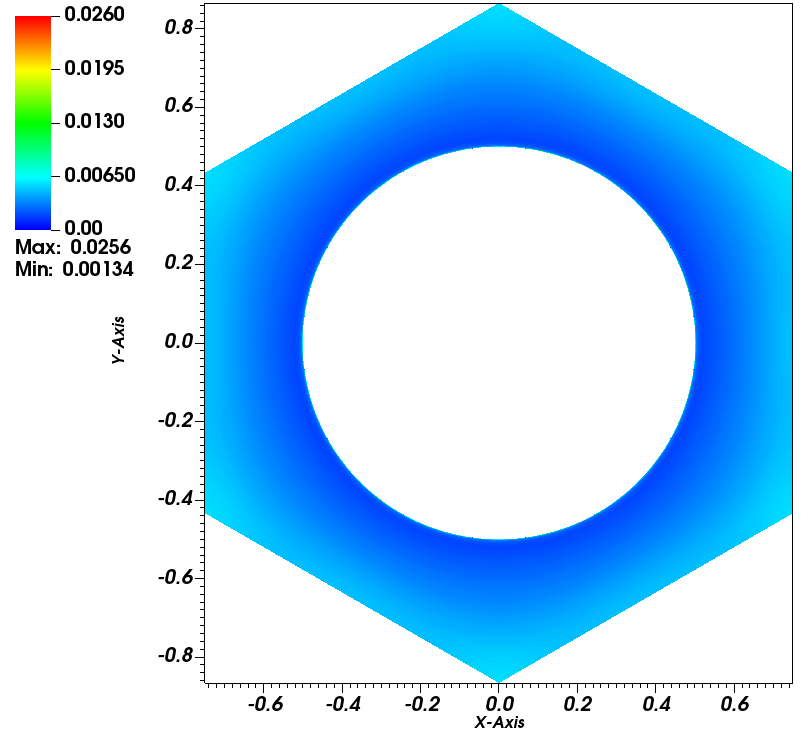} }}%
    \caption{DNS mesh size (left) compared with the Kolmogorov length scale obtained with RANS (right)}%
    \label{fig4}%
\end{figure}

The mesh was considered suitable for DNS after the comparison between the mesh size with the Kolmogorov length scale, as shown in Figure \ref{fig4}, yielded values below the Kolmogorov length scale. That ensured that the mesh could resolve the smallest eddies everywhere. 
\par

\subsection{Comparison of  RANS and LES results}

As discussed in the previous section, we perform URANS simulations with a Reynolds number of 10000 and a coarse mesh with 58800 hexahedral elements. Despite the low Reynolds number, no low-Reynolds damping is applied. The operator integrator splitting factor (OIFS) was implemented to allow the simulation to run at larger Courant numbers, up to 3.5, and the time step size of $dt = 10^{-3}D/u$.

\begin{figure}[h]%
    \centering
    \subfloat \centering  \includegraphics[width=8.2cm]{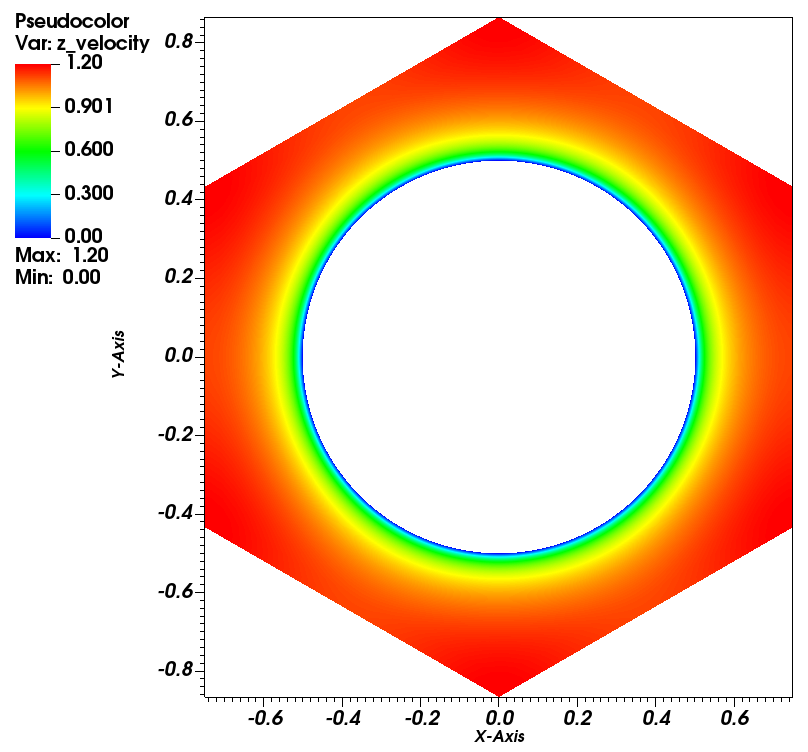}
    \subfloat \centering \includegraphics[width=8.2cm]{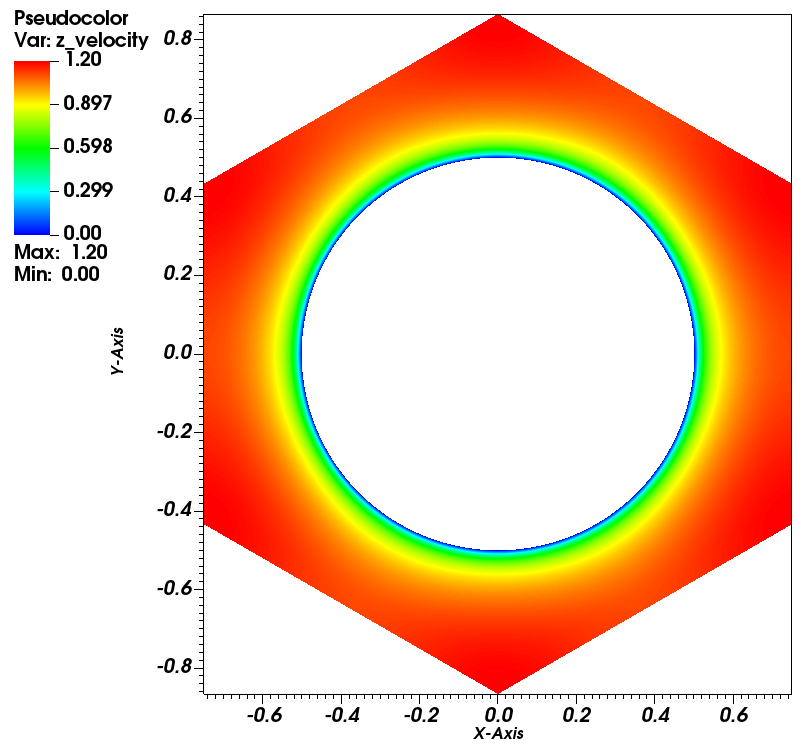}
    \caption{Streamwise velocity for RANS with polynomial order 4 and 10, respectively}%
    \label{fig5}%
\end{figure}

\par
\begin{figure}[H]
  \centering
     \includegraphics[width=0.63\textwidth]{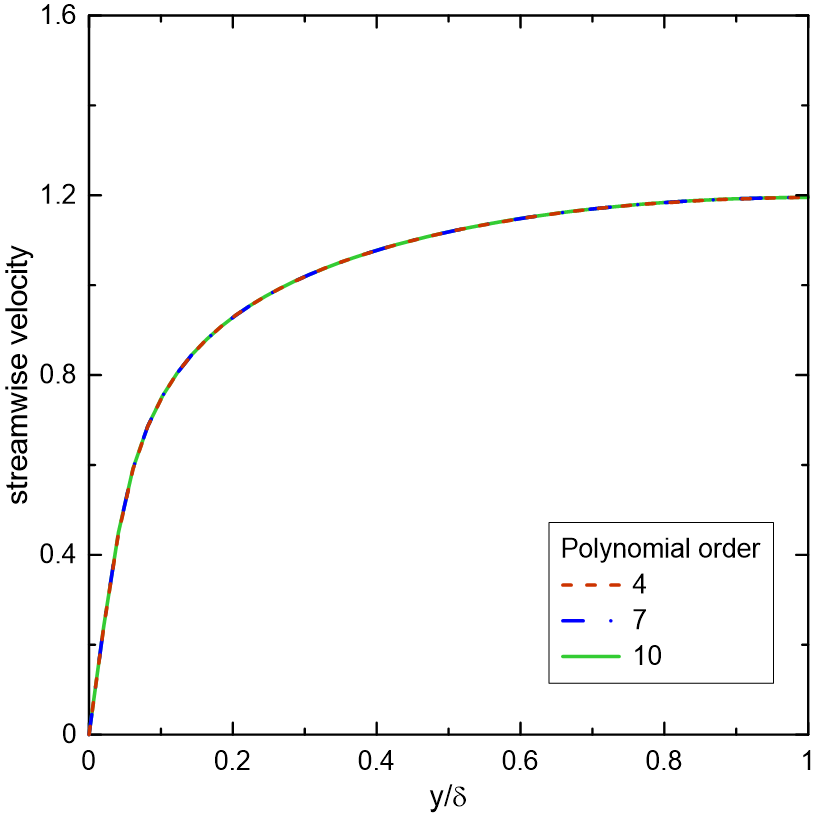}
  \caption{Comparison between RANS simulations at different polynomial orders. Streamwise velocity  as a function of the normalized wall distance in the wide gap}
  \label{fig6}
\end{figure}

 Three cases of RANS calculations with the $k-\tau$ model and the Gauss-Lobatto Legendre scale, $P_{N+1}$, of 5, 8, and 11 were compared, as represented in Figures \ref{fig5} and Figures \ref{fig6}. 
\par The streamwise velocity for the lower and the higher polynomial orders was in excellent agreement with each other, demonstrating the RANS results' quality.  Therefore, the lower order was chosen to conduct the RANS simulations, which saved computational cost for all the turbulence models tested.
\vspace{16pt}
\par
\begin{figure}[H]%
    \subfloat \centering  {{\includegraphics[width=7.95cm]{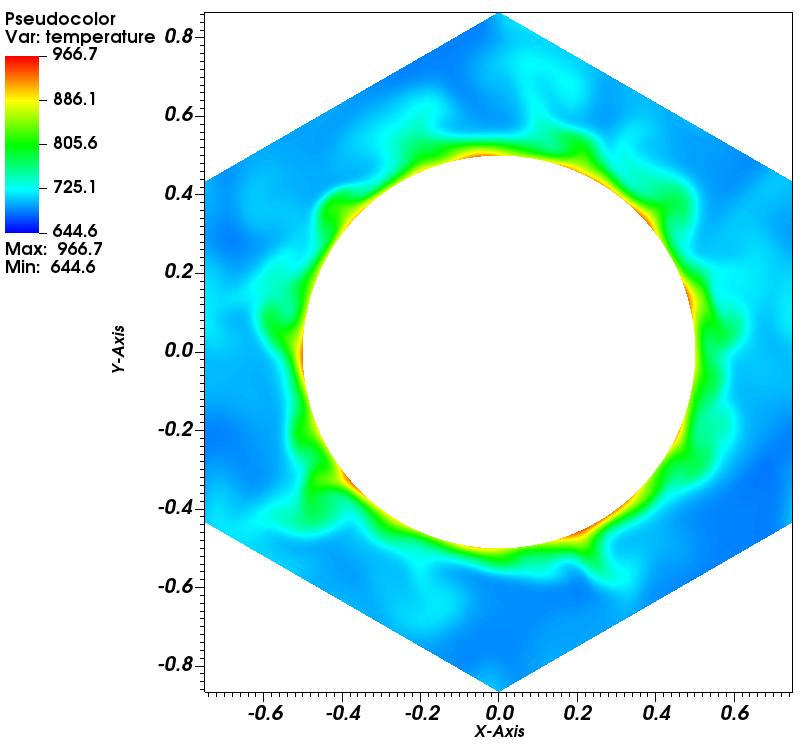} }}%
    \subfloat \centering {{\includegraphics[width=7.95cm]{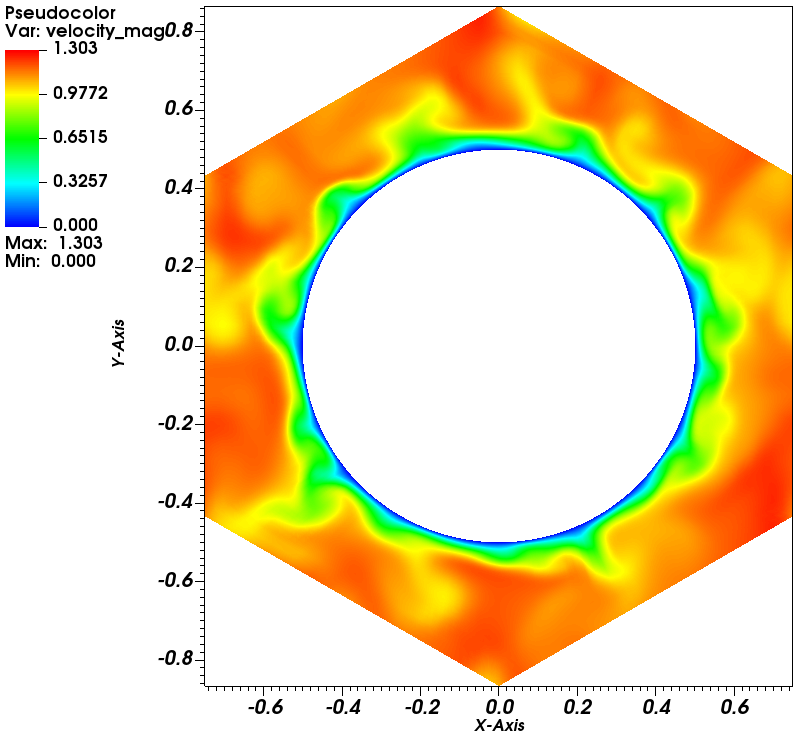} }}%
    \caption{Cross section of temperature and velocity fields for LES , respectively}%
    \label{fig7}%
\end{figure}

\begin{figure}[H]%
    \centering
    \subfloat \centering  {{\includegraphics[width=13cm]{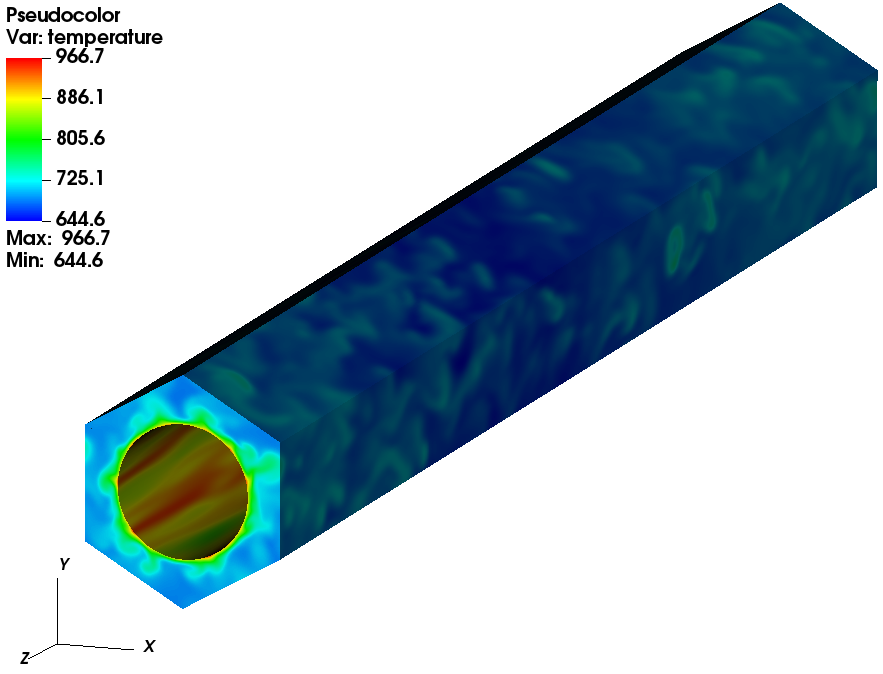} }}%
    \subfloat \centering {{\includegraphics[width=13cm]{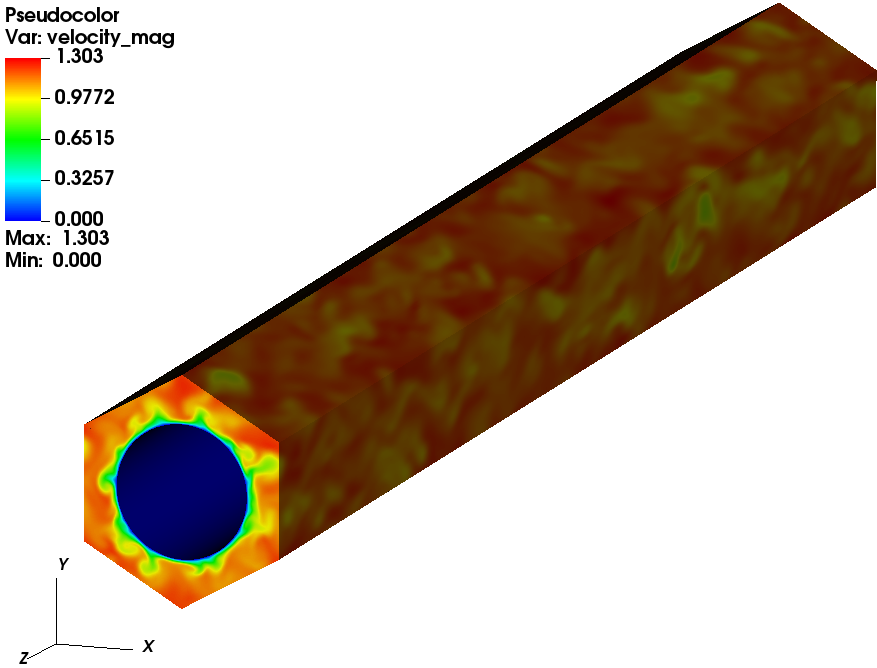} }}%
    \caption{Temperature and velocity fields for LES, respectively}%
    \label{fig8}%
\end{figure}

 In addition to RANS, we performed a series of wall-resolved LES to verify the results obtained. The LES calculations were conducted on a finer mesh, designed to resolve Taylor microscale, with a time step size of $dt = 10^{-3}D/u$. A second-order time-stepping scheme was used with a Courant number below 0.4. The cross-section and the whole domain for the temperature and velocity fields are shown in Figures \ref{fig7} and \ref{fig8}, where the flow structures can be observed.


\begin{figure}[h]%
    \centering
    \subfloat \centering  {{\includegraphics[width=7.95cm]{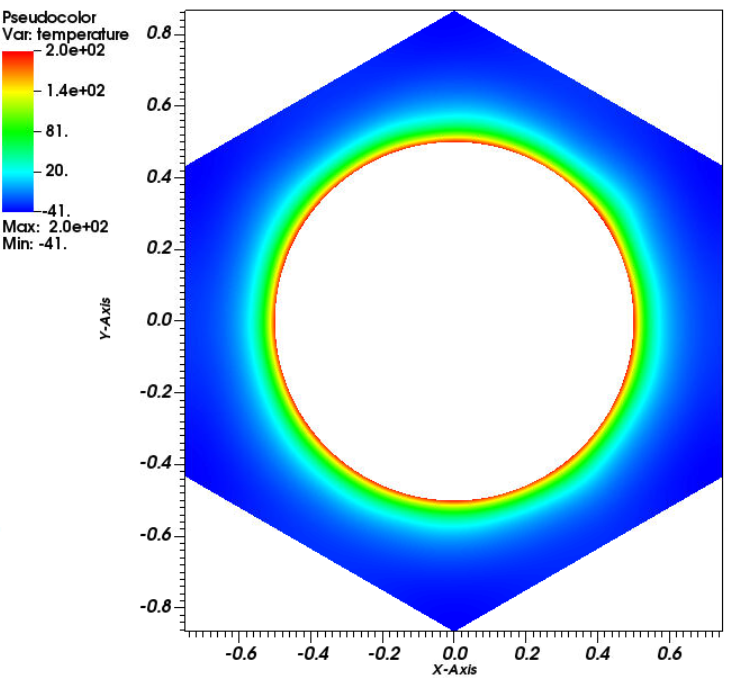} }}%
    \subfloat \centering {{\includegraphics[width=7.95cm]{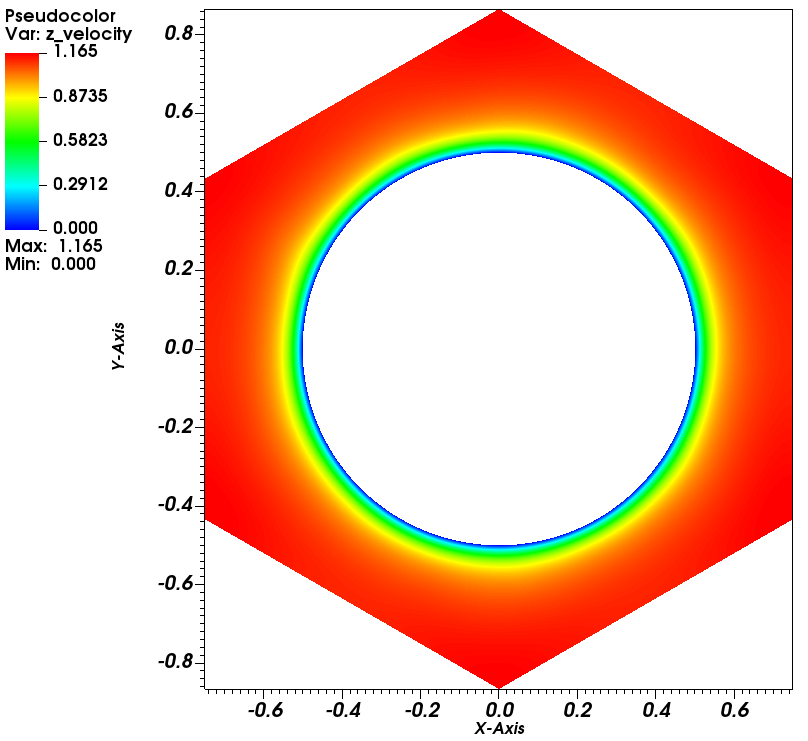} }}%
    \caption{Average temperature and average streamwise velocity, respectively}%
    \label{fig9}%
\end{figure}

After the initial transient passed and the statistical steady state was achieved, the LES results for the streamwise velocity and temperature fields were time-averaged, and planar averaged using 30 folders with 30 $D/u$ in each, as represented in Figure \ref{fig9}. 

\begin{figure}[H]
  \centering
     \includegraphics[width=0.56 \textwidth]{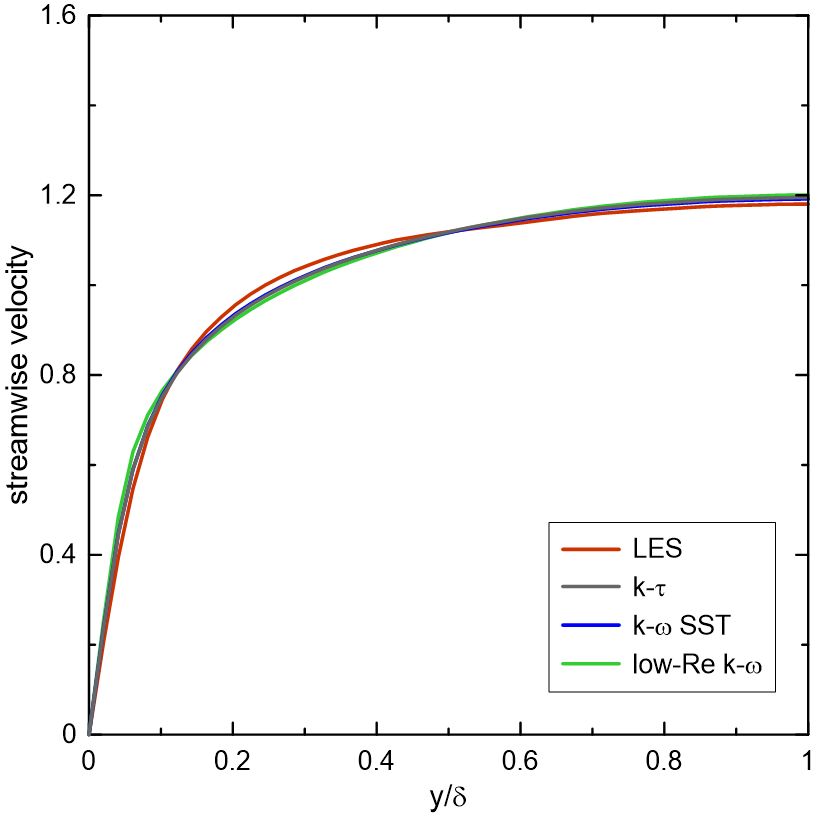}
  \caption{Comparison between RANS streamwise velocity and LES streamwise average velocity as a function of the normalized wall distance in the wide gap}
  \label{fig10}
\end{figure}

The averaged streamwise velocity field, in Figure \ref{fig9} and RANS streamwise velocity field, in Figure \ref{fig5} presented similar results, qualitatively. A comparison of those two variables, in a line along the wide gap, as shown in Figure \ref{fig10}. The RANS streamwise velocity was computed for three turbulence models: $k-\tau$, $k-\omega$  $SST$, and $low-Re$ $k-\omega$. The simulations demonstrate a remarkable degree of agreement for a rod bundle flow, likely a factor of the much higher P/D and the comparison location in the wide gap.

\begin{figure}[H]%
    \centering
 {{\includegraphics[width=9.5cm]{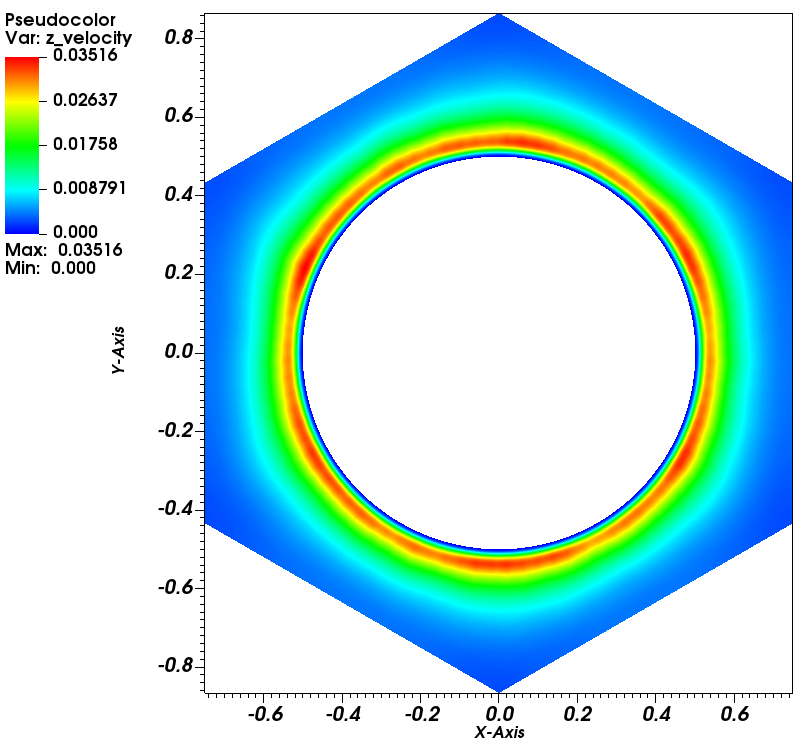} }}%
    \caption{RMS streamwise velocity}%
    \label{fig11}%
\end{figure}

In addition, the cross section of the RMS of the temperature and streamwise velocity fields were represented in Figure \ref{fig11}. 

\begin{figure}[H]%
    \centering
    \subfloat \centering  {{\includegraphics[width=8.2cm]{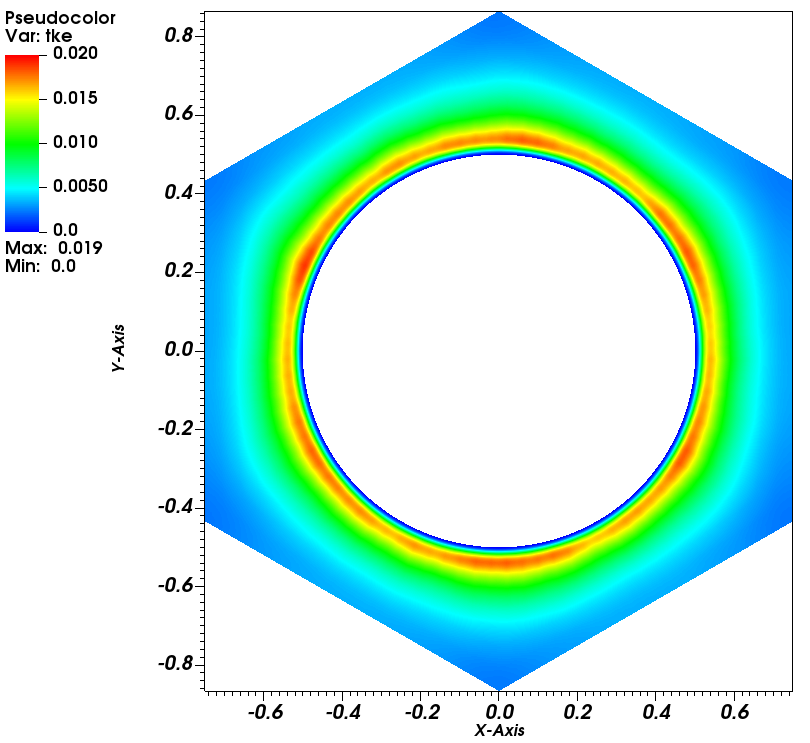} }}%
    \subfloat \centering {{\includegraphics[width=8.2cm]{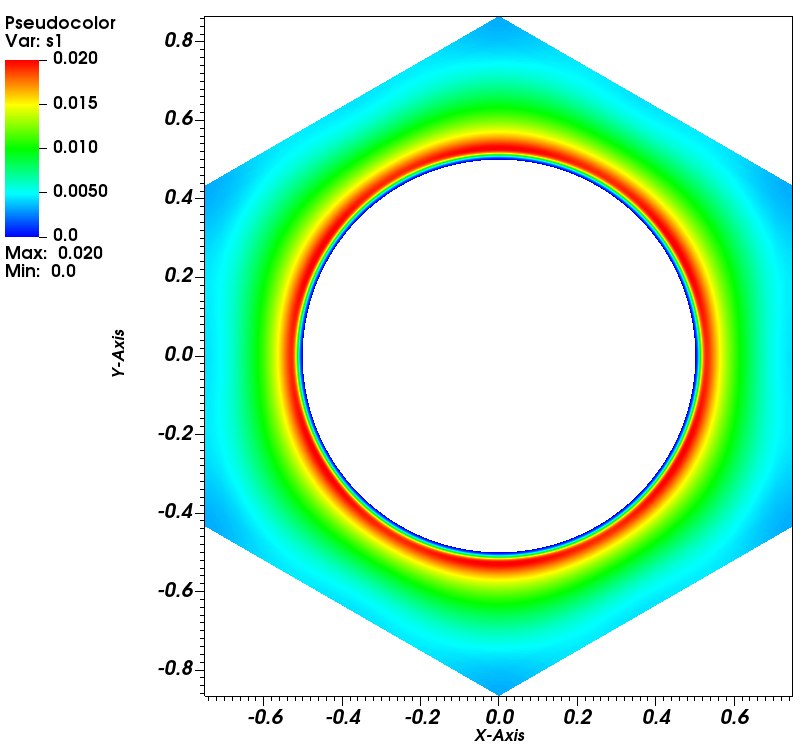} }}%
    \caption{Turbulent Kinetic Energy for LES and $low-Re$ $k-\omega$}%
    \label{fig12}%
\end{figure}

\begin{figure}[H]
  \centering
     \includegraphics[width=0.6 \textwidth]{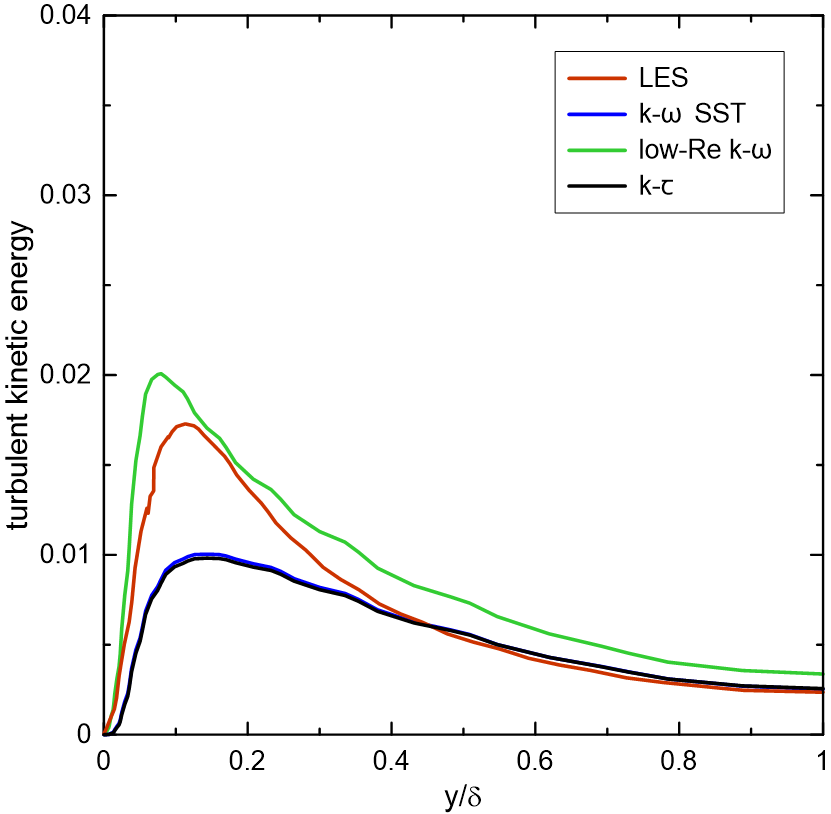}
  \caption{Turbulent Kinetic Energy as a function of the normalized wall distance in the wide gap}
  \label{fig13}
\end{figure}

Then, the turbulent kinetic energy $k$ was obtained from the RANS models and calculated for the LES, using Equation \ref{eq9}. The results in a cross-section of the domain are represented in Figure \ref{fig12}. Besides, this variable was computed in a line along the wide gap, as shown in Figure \ref{fig13}.


The LES and the low-Re $k-\omega$ model results presented good agreement, especially in the near-wall region. On the other hand, the profiles for the other two turbulence models diverge from the LES as expected. This occurred due to the lack of damping in the RANS wall models, in general, due to the simplified turbulence models used. The damping functions present in low-Reynolds models ensure that effects caused by the viscous stresses near the wall are more important than the turbulent Reynolds' stresses \cite{JOSHI201921}.

Lastly, we computed the Nusselt number through its definition:
\begin{equation}
    Nu = \frac{h D_h}{\lambda}
\end{equation}
where $h$ is the convective heat transfer coefficient, given by:
\begin{equation}
    h = q/(T_{wall}-T_{bulk})
\end{equation}
where $T_{wall}$ is the average temperature in the wall, $T_{bulk}$ is the fluid bulk temperature, and $q$ is the heat flux.
And then we compared the value obtained with the Gnielinski correlation\cite{Gnielinski1975}, valid for tubes in the range $3000 \leq  Re_{D_h} \leq  5 \times 10^6$ and $0.5 \leq Pr \leq  2000$:

\begin{equation}
    Nu_{D_h} = \frac{(f/8)(Re_{D_h}-1000)Pr}{1+12.7(f/8)^1/2(Pr^2/3 -1)}
\end{equation}
where $f$ is the Darcy friction factor, obtained through the wall shear stress $\tau_{\omega}$.
\par The results, summarized in Table \ref{table1}, presented good agreement even thought the Gnielinski correlation is more general and suitable for smooth pipe flows.
\vspace{16pt}
\begin{table}[!h]
\centering
\caption{Comparison between Gnielinski correlation and the definition of Nusselt number}
\label{table1}
\vspace{14pt}
\begin{tabular}{||r||c||} \hline \hline
        $Nu$ &  35.4\\ \hline
        $Nu_{D_h}$ & 37.2\\ \hline
        Relative Error& 5\%\\ \hline \hline
\end{tabular}
\end{table}
\vspace{16pt}

\section{Conclusions}

Complementing experimental databases, high fidelity CFD simulations are crucial for insight into
nuclear reactor core components and improving their safety. Although RANS simulations are widely used in
engineering applications, they are limited in predicting the coolant flow with rod bundles, as they can only
provide an approximate description of turbulence. Therefore, we aim to perform DNS calculations in
a subchannel to provide data to build closure models for advanced reactors.
\par This work lays the groundwork to define a numerical DNS dataset for helium-cooled hexagonal rod
bundles. We used RANS simulations to predict the Kolmogorov length scale and obtain a mesh suitable to
perform DNS, which will be the next step of this work.
\par Furthermore, we performed preliminary LES calculations to compare with the RANS models. The velocity
field was time-averaged, and planar averaged, providing results that agreed well with the one obtained with
the RANS simulations. The turbulent kinetic energy results obtained with LES were also in reasonable
agreement with the RANS results, building additional confidence in the RANS simulations used to
estimate the Kolmogorov length scale. The planar average provided reliable LES results, in
agreement with the low-Reynolds RANS model tested in this work.

\section*{Acknowledgments}
The authors appreciate the support of the U.S. Department of Energy, Office of Nuclear Energy, the federal and technical points of contact, and the external review committee.

\setlength{\baselineskip}{12pt}

\bibliographystyle{nureth18new}
\bibliography{references}

\end{document}